\documentclass[usenatbib]{aa}
\usepackage{times}
\usepackage{graphicx}
\usepackage{natbib}
\usepackage{subfigure}
\usepackage{dcolumn}
\usepackage{longtable}
\usepackage{lscape}
\usepackage {ulem} 
\usepackage[usenames,dvipsnames]{xcolor}
\usepackage{xspace}
\usepackage{aas_macros}
\usepackage{xargs}
\usepackage{ulem}
\usepackage{url}

\newcommand{\galfit}{{\scshape galfit}\xspace}

\newcommand{\galfitm}{{\scshape galfitm}\xspace}

\newcommand{\megamorph}{{MegaMorph}\xspace}
\newcommand{\sersic}{S\'ersic\xspace}

\newcommand{\ferengi}{{\scshape ferengi}\xspace}

\newcommandx{\N}[2][1= ,2= ]{$\mathcal{N}^{#1}_{#2}$\xspace}
\newcommandx{\R}[2][1= ,2= ]{$\mathcal{R}^{#1}_{#2}$\xspace}
\newcommand{\re}{R_{\rm e}}

\newcolumntype{d}{D{.}{.}{-1}}

\title{MegaMorph: classifying galaxy morphology using multi-wavelength \sersic profile fits}
\author{Marina~Vika\inst{\ref{inst1}}$^,$\inst{\ref{inst2}}$^,$\inst{\ref{inst3}}
\and Benedetta~Vulcani\inst{\ref{inst4}} 
\and Steven~P.~Bamford\inst{\ref{inst5}} 
\and Boris~H{\"a}u{\ss}ler\inst{\ref{inst6}}$^,$\inst{\ref{inst7}} 
\and Alex~L.~Rojas\inst{\ref{inst2},\ref{inst8}}
}

\institute{Institut f{\"u}r Astr- und Teilchenphysik, Universit{\"a}t Innsbruck, Technikerstra{\ss}e 25, A-6020 Innsbruck, Austria \email{vika.marina@gmail.com} \label{inst1}
\and Carnegie Mellon University in Qatar, Education City, PO Box 24866, Doha, Qatar\label{inst2} 
\and Institute for Astronomy, Astrophysics, Space Applications \& Remote Sensing, National Observatory of Athens, P. Penteli, 15236, Athens, Greece\label{inst3} 
\and Kavli Institute for the Physics and Mathematics of the Universe (WPI), Todai Institutes for Advanced Study, University of Tokyo, Kashiwa 277-8582, Japan\label{inst4}
\and School of Physics and Astronomy, The University of Nottingham, University Park, Nottingham, NG7 2RD, UK\label{inst5} 
\and Department of Physics, University of Oxford, Denys Wilkinson Building, Keble Road, Oxford, OX1 3RH, UK\label{inst6} 
\and University of Hertfordshire, Hatfield, Hertfordshire, AL10 9AB, UK\label{inst7}
\and Carnegie Mellon University, 5000 Forbes Ave, Pittsburgh, PA 15213, US\label{inst8} 
}

\begin{document}

\date{Received .../ 
Accepted ...}

\label{firstpage}

\abstract{}
{This work investigates the potential of using the wavelength-dependence of galaxy structural parameters (\sersic index, $n$, and effective radius, $R_e$) to separate galaxies into distinct types.
}
{A sample of nearby galaxies with reliable visual morphologies is considered, for which we measure structural parameters by fitting multi-wavelength single-\sersic models. Additionally, we  use a set of artificially redshifted galaxies to test how these classifiers behave when the signal-to-noise decreases.
}
{We show that the wavelength-dependence of $n$ may be employed to separate visually-classified early- and late-type galaxies, in a manner similar to the use of colour and $n$. 
Furthermore, we find that the wavelength variation of $n$ can recover galaxies that are misclassified by these other morphological proxies.
Roughly half of the spiral galaxies that contaminate an early-type sample selected using $(u-r)$ versus $n$ can be correctly identified as late-types by \N, the ratio of $n$ measured in two different bands.
Using a set of artificially-redshifted images, we show that this technique remains effective up to $z\sim 0.1$. 
\N can therefore be used to achieve purer samples of early-types and more complete samples of late-types than using a colour-$n$ cut alone.
We also study the suitability of \R, the ratio of $R_e$ in two different bands, as a morphological classifier, but find that the average sizes of both early- and late-type galaxies do not change substantially over optical wavelengths.
}{}

\keywords{
galaxies: photometry --
galaxies: elliptical and lenticular, cD --
galaxies: spiral --
galaxies: fundamental parameters --
galaxies: structure ---
techniques: image processing
}

\titlerunning{MegaMorph: galaxy morphologies}
\authorrunning{Vika et al.}
\maketitle

\section{Introduction}
\label{sec:intro}


The visual morphology of galaxies, provides important information regarding their star-formation and dynamical histories, without requiring expensive spectroscopy.
In particular, the variation of morphology with mass and environment have provided key insights into galaxy evolution (\citealt{tex:D80, tex:BN09}).
However, visual classification is only possible for samples with sufficient spatial resolution, requires a huge amount of effort for large samples, and is somewhat subjective \citep[and references therein]{tex:Sa05,tex:LS11}. 
Galaxy morphology is broadly correlated with many other -- more objective -- observables, such as luminosity, colour, structure and light profile shape. 
Galaxies may therefore be morphologically classified using various approximate techniques (e.g. \citealt{tex:SI01, tex:VW08, tex:CF11}). 


A variety of parametric and non-parametric methods are generally being used for classifying galaxies.
For instance, one way of characterising  morphologies is by fitting the light profile of a galaxy with analytic functions and measuring parameters, such as the \sersic index \citep{tex:S68}.
Non-parametric quantities, such as concentration index, colour, Concentration-Asymmetry-Clumpiness parametrisation, and Gini-$M_{20}$ coefficients, are also commonly used \citep{tex:AT96,tex:SF01,tex:C03,tex:LP04,tex:PC05,tex:SC07b}.
Each of these techniques has  advantages and disadvantages.
The main benefit of non-parametric methods is that they do not rely on prior assumptions about galaxy structure, although they generally do need some prior knowledge (e.g. the galaxy centre, masking, etc.). 
They also require calibration if one wishes to interpret them in terms of standard visual morphologies.
Even then, these methods typically depend on image depth and resolution; different datasets for the same galaxy can give opposing results.
In contrast, \sersic fitting assumes that each galaxy has a light distribution that can be well described by a \sersic function, but is generally more stable to varying image quality \citep{tex:HM07}.


In large surveys it has become popular to fit single-\sersic profiles to all galaxies, in order to  recover total magnitudes, effective radii ($\re$), and obtain morphological information from the \sersic indices ($n$) (e.g. \citealt{tex:vB12}).  
Using these products, a division in the optical (or optical $-$ near-infrared) colour versus $n$ plane is often applied in an attempt to separate out early- and late-type galaxies,  i.e. dividing into red, high-$n$ and blue, low-$n$ galaxies (e.g. \citealt{tex:KD12}).  
This technique is successful for the majority of galaxies, however it has some shortcomings. 
In particular, many spirals (Sa/Sb) are found to inhabit the same region of the colour versus \sersic index plane as ellipticals and lenticulars (\citealt{tex:VB14b}).


We have recently developed and demonstrated a technique for fitting wavelength-dependent \sersic models to galaxy surface-brightness profiles in a simultaneous and consistent manner (\citealt{tex:BH12b}, \citealt{tex:HB13},  \citealt[hereafter V13]{tex:VB13b}). 
This technique provides measurements of the wavelength variation of \sersic index (i.e., the ratio of $n$ at particular pairs of wavelengths, denoted \N) and effective radius (i.e., the ratio of $\re$ at particular pairs of wavelengths, denoted \R) for individual galaxies. 
\citet[hearafter V14]{tex:VB14} showed that `early-type' (red with $n_r>2.5$) and `late-type' (blue with $n_r<2.5$) galaxies display different \N and \R distributions. 
The \sersic index is found to vary significantly with wavelength for late-types, but to remain constant for early-types.  
This is possibly an effect of late-types possessing a bluer disk, which contrasts with a redder bulge, while early-types are dominated by a spheroid at all wavelengths, and/or their bulge and disk components have similarly red colours. 
A detailed analysis of the component colours can be found in \citet{tex:VB14b} where we present the results of the bulge-disk decomposition of a sample of 163 nearby galaxies. 


As far as $R_e$ is concerned, V14 found that it  becomes smaller with increasing wavelength for both early- and late-type galaxies, with the effect of being larger for early-types. 
The wavelength dependence of effective radius is weaker in optical-only data compared to the optical-NIR data used in V14.  
Therefore, \R requires NIR data to be useful as a morphological classifier.
However, V14 found that \R is less effective than \N at separating early- and late-types.
In general, \N and \R provide a simple, but powerful, parametric way of characterising galaxy colour gradients, which reveal information about the galaxy internal structure and its relation to other galaxy properties.
A shortcoming of the analysis in V14 is that the morphologies of the galaxies in that sample are unknown. 
So while it gives clues and shows trends, an actual comparison on how well these values can be used to reproduce visual morphological classifications, is impossible. 
This is what we investigate in this paper.

At a more practical level, quantifying how $n$ and $\re$ vary with wavelength is crucial for removing biases when comparing measurements made using different bandpasses or at different redshifts.
Also, it makes comparisons between different datasets or between different redshifts both easier and more consistent.

The aim of this paper is to prove the correspondence between $n$, \N, \R and morphology. 
We will show that the wavelength dependence of \sersic index and effective radius can be used to separate the majority of visually-classified  ellipticals from lenticulars and spiral galaxies, in a manner similar to commonly applied methods using colour and \sersic index.


\begin{table*}
\caption{The ability of the two classification methods presented in top right- and left-panels of Figure~\ref{fig:MMzg} to separate elliptical galaxies.
}
  \smallskip
  \centering
  \begin{tabular}{cccccccc}
  \hline
  \noalign{\smallskip}
Morphology & Total& $\left<{\rm n}_r\right>$ &  $\left<{\rm B/T}_r\right>$ & (u-r)$>2.3$    & (u-r)$>2.3$    &  (u-r)$_{0.11}>2.3$ & (u-r)$_{0.11}>2.3$   \\  
           &      &  & & and n$_r>2.5$  & and $$\N[z][g]$<1.2$ &   and n$_{r,0.11}>2.5$     & and $$\N[z][g,0.11]$<1.2$    \\ 
\noalign{\smallskip}
\hline
 \noalign{\smallskip}
\hline
 \noalign{\smallskip}
E       & 19      & 4.1 & 0.67 &16     & 18     & 12    & 14     \\
S0      & 13 (2)  & 3.3 & 0.62 &10 (1) &  5 (1) & 7 (0) &  8 (1) \\
Sa      & 2  (1)  & 4.4 & 0.45 & 2 (1) &  1 (0) & 1 (0) &  1 (0) \\ 
Sab-Sb  & 27 (17) & 2.9 & 0.32 & 9 (5) &  3 (1) & 6 (2) &  5 (2) \\  
Sbc-Sc  & 49 (32) & 1.5 & 0.21 & 1 (1) &  0     & 0     &  0     \\ 
Scd-Sd  & 21 (16) & 1.2 & 0.24 & 0     &  0     & 0     &  0     \\  
Sm-Irr  & 11 (9)  & 1.2 & 0.11 & 0     &  0     & 0     &  0     \\
\noalign{\smallskip}
\hline
\end{tabular}
\tablefoot{Column 1: Hubble type, Column 2: Total number of galaxies in each Hubble type after applying the following cuts: $M_{r}<-18.7$ and $z<0.01$, Columns 3: Mean n$_r$ from V13, Column 4: Mean B/T from \citet{tex:VB14b}, Columns 5-6: Number of galaxies selected in each cut for the original sample. Columns 7-8: Number of galaxies selected in each cut at redshift 0.11. The number of barred galaxies are shown in brackets. Where \N[z][g] is the ratio of $n$ measured in z- and g- bands ($n_z/n_g$).
}
\label{table:sample} 
\end{table*}

\begin{table*}
\caption{ The number of galaxies for each Hubble bin for the remaining three quadrants of the top right- and left-panels of Figure~\ref{fig:MMzg}.
}
  \smallskip
  \centering
  \begin{tabular}{cccccccc}
  \hline
  \noalign{\smallskip}
Morphology & Total  &  (u-r)$>2.3$   &  (u-r)$>2.3$ & (u-r)$<2.3$  & (u-r)$<2.3$  & (u-r)$<2.3$  & (u-r)$<2.3$     \\  
           &       & and n$_{r}<2.5$ & and $$\N[z][g]$>1.2$ & and n$_{r}<2.5$ & and $$\N[z][g]$>1.2$  & and n$_{r}>2.5$  & and $$\N[z][g]$<1.2$   \\ 
\noalign{\smallskip}
\hline
 \noalign{\smallskip}
\hline
 \noalign{\smallskip}
E       & 19   &  3   & 1  & 0  &  0 & 0  & 0      \\
S0      & 13   &  2   & 7  & 0  &  0 & 1  & 1      \\
Sa      & 2    &  0   & 1  & 0  &  0 & 0  & 0      \\ 
Sab-Sb  & 27   &  5   & 11 & 8  & 10 & 5  & 3      \\  
Sbc-Sc  & 49   &  3   & 4  & 41 & 36 & 4  & 9      \\ 
Scd-Sd  & 21   &  0   & 0  & 21 & 19 & 0  & 2      \\  
Sm-Irr  & 11   &  0   & 0  & 11 &  7 & 0  & 4      \\
\noalign{\smallskip}
\hline
\end{tabular}
\tablefoot{Column 1: Hubble type, Column 2: Total number of galaxies in each Hubble type after applying the following cuts: $M_{r}<-18.7$ and $z<0.01$, Columns 3-8: Number of galaxies selected in each cut for the original sample.}
\label{table:sample2} 
\end{table*}

\section{Data}
\label{sec:data}

We use  a small, but well-characterised sample of galaxies at $z<0.01$ ({\it original sample}) and a secondary sample of artificially-redshifted images ({\it redshifted sample}). 
The latter allows us to assess the accuracy and reliability of galaxy profile fitting over a wide range of spatial resolution and signal-to-noise. 
These samples are presented  in V13, where full details regarding the sample selection, the redshifting procedure and the fitting process for both original and redshifted images can be found. 
Briefly, the original sample consists of 163 nearby galaxies of various morphologies (taken from NED\footnote{NASA/IPAC Extragalactic Database; http://ned.ipac.caltech.edu/}) with images in $u$, $g$, $r$, $i$, $z$ bands.
All the morphological  classes, except for ellipticals, also include barred galaxy types. 
These galaxies were chosen because they have parametric surface-brightness profile measurements by previous studies (using single-band data, see  \citealt{tex:MG98,tex:CC93,tex:M04,tex:PT06}), and have available imaging data from SDSS \citep{tex:AA09}.

To construct the redshifted sample we used the \ferengi software \citep{tex:BJ08} to apply cosmological changes in angular size and surface brightness, in order to simulate observations at greater distances.  
We simulated images at redshifts $0.01$--$0.25$, in steps of $0.01$ without applying a k-correction. 
In this way the redshifted galaxies have a constant absolute magnitude in each band as a function of redshift. 
This is less physically realistic, but allowed us in V13 to test the ability of \galfit to measure magnitudes as the noise increases and resolution decreases.
Throughout the paper we use the term `artificially-redshifted' to refer to these images.

Colours, $n$, \N and \R values were measured by fitting single-\sersic profiles using \galfitm (Bamford et al., in prep.). 
\galfitm\footnote{\galfitm is publicly available at \url{http://www.nottingham.ac.uk/astronomy/megamorph/}.} fits a single wavelength-dependent model to all the provided images simultaneously. 
Rather than fitting the parameter values at the wavelength of each band, \galfitm fits the coefficients of a smooth function describing the wavelength dependence of each parameter. 
For this paper, we allow full freedom in magnitudes, while \sersic index and $\re$ are allowed to vary linearly with wavelength.
All other parameters (center position, axis ratio and position angle) are selected to be constant with wavelength, e.g. the same in all images. 
Constraining the size and shape of the galaxy in this manner improves the robustness of the measured colours and other parameters.
However, while these constraints increase the stability of the fits and reduce the statistical uncertainties of the structural measurements, there is also the risk of introducing systematic biases in cases where the true wavelength dependence of the profile does not correspond to that assumed. 
Such systematics may be reduced by giving the model more freedom to vary with wavelength, at the cost of increased statistical uncertainties on the measured parameters in the low signal-to-noise bands.
For a more detailed discussion on the benefits and potential biases of fits with linearly constrained \sersic index and $\re$ see V13.

To ensure completeness in the original sample, we only select galaxies that are brighter than $M_r = -18.7$~mag and have $z<0.01$. 
After applying these restrictions, we retain a sample of 142 galaxies. 
The number of galaxies for each Hubble type is given in Table 1. Our sample contains 77 galaxies with a bar ($\sim$54\%).

\begin{figure*}
\centering
9\includegraphics[width=0.75\textwidth]{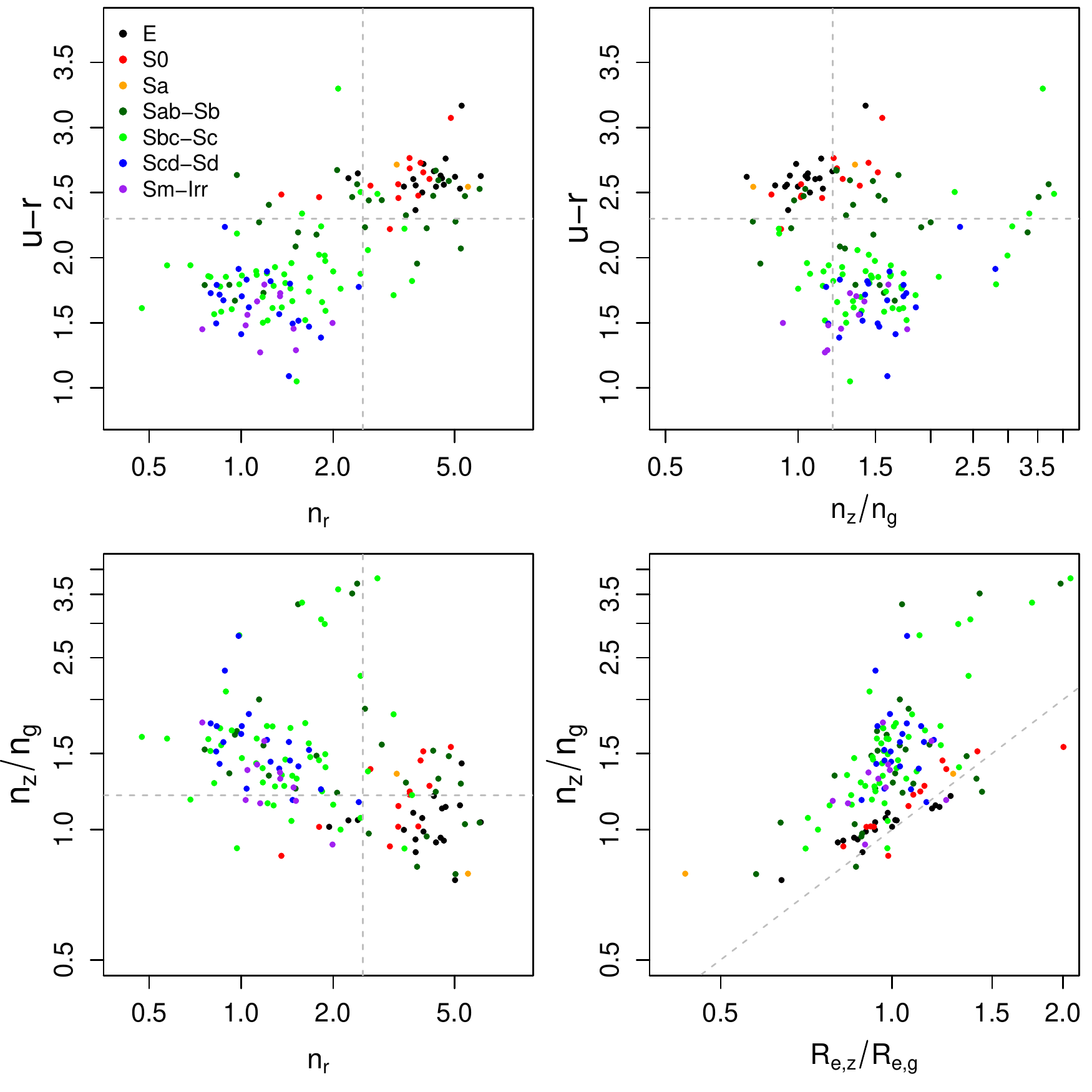}
\caption{Original images. Top-left panel: $u-r$ colour versus the \sersic index in r-band.
Top-right panel: $u-r$ colour versus \N[z][g]. 
Bottom-left panel: \N[z][g]  versus \R[z][g]. 
Bottom-right panel: \N[z][g] versus \R[z][g].
All the parameters have been derived with multi-band fitting with the use of \galfitm.  
The dashed lines are in the following values: $u-r=2.3$, $n_r=2.5$ and \N[z][g]~$= 1.2$. In the bottom right panel we plot the 1:1 line. 
}
\label{fig:MMzg}
\end{figure*}

\begin{figure*}
\centering
\includegraphics[width=0.75\textwidth]{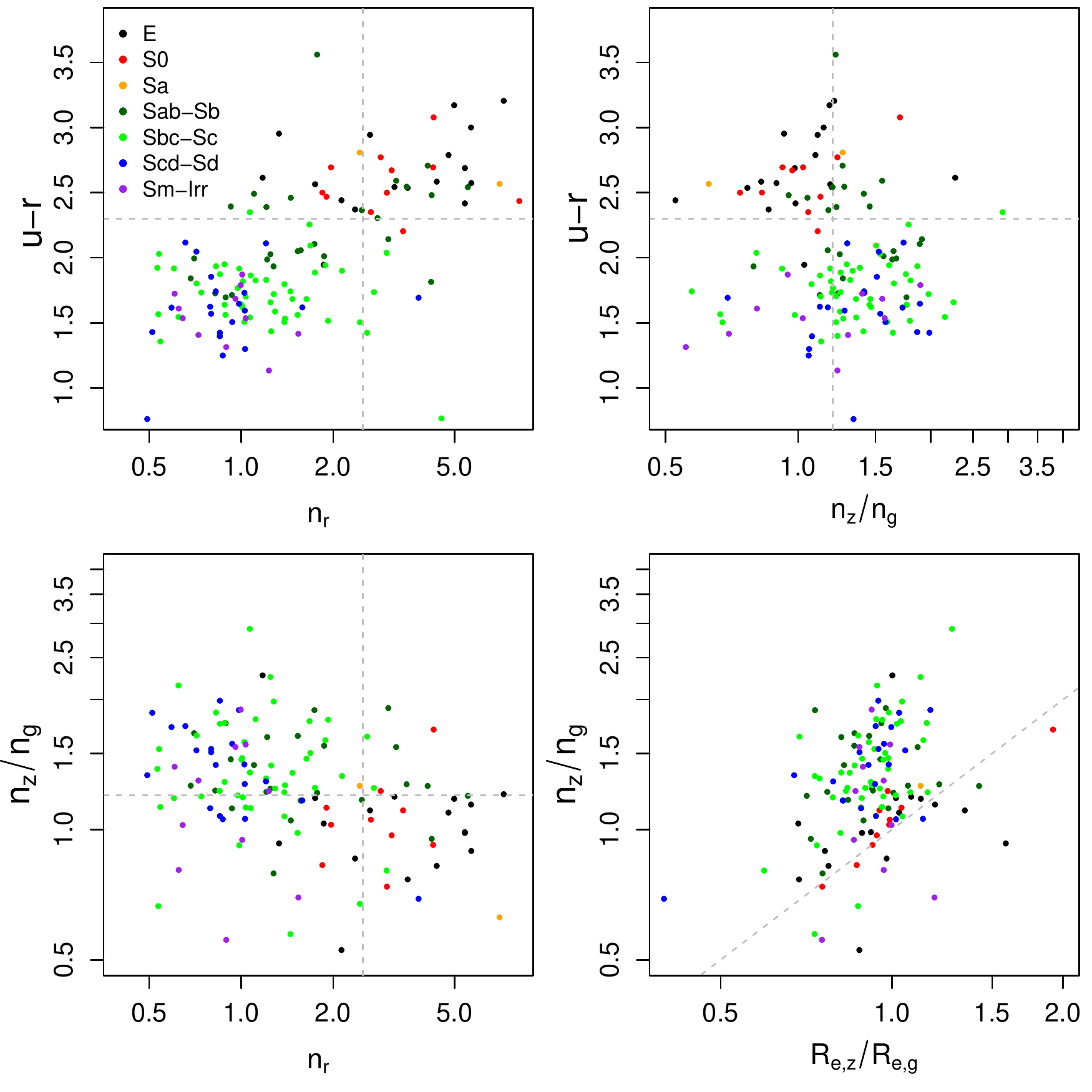}
\caption{Same as Fig.~\ref{fig:MMzg} but for artificially redshifted galaxies at $z=0.11$.}
\label{fig:MMz}
\end{figure*}

\section{Results \& Discussion}
\label{sec:results}

In Figure~\ref{fig:MMzg} we present four ways of using our automated measurements to separate visually-classified early- and late-type galaxies for the original sample.  
Table~\ref{table:sample} gives the number of galaxies of each Hubble type selected as `early-types' by each set of classification cuts. 
 For comparison, we also include the average \sersic index and bulge-fraction (from \citealt{tex:VB14b}) for each morphology.
 Table~\ref{table:sample2} tabulates the number of galaxies in the remaining three quadrants.
We select the cuts by eye, in order to maximise the overlap between the two classification methods.


The first panel illustrates one of the most commonly used ways of separating late-type  from early-type galaxies in large surveys, where the morphological classification of the galaxies is unknown.
The vast majority of the later type  galaxies (from Sbc to Irr) are found in the so-called `blue cloud', having $(u-r)<2.3$ and  $n_r<$2.5, with all of the Scd-Irr's  falling in this section of this diagram.
In contrast, most of the E, S0 and Sa galaxies are located in the `red cloud' 
($(u-r)>2.3$,  $n_r>$2.5), but some Sbs also fall into this segment. 
Between these  two main clouds we find most of the Sab-Sb galaxies in our sample. 
It is interesting to note that there are no Es and S0s in the blue cloud, with the few E-S0 galaxies with $n_{r}<2.5$ being redder than the equivalent late-type galaxies. 
The fact that visually-classified early and late types are separated in this plane confirms the correspondence connection between colour, $n$ and morphology. 

Most authors in previous studies have used either a \sersic index or  a colour cut to separate early-type from late-type galaxies (e.g. \citealt{tex:RF04}). 
However, as can be seen in this first panel, neither of these simple cuts defines clean and complete samples of galaxies.
A combination of both parameters produces somewhat cleaner samples (see, e.g., \citealt{tex:KD12}).


The upper right panel presents a new way of classifying galaxies. 
This figure was introduced in V14 where we discussed the possibility to use it to identify very special populations of galaxies, e.g. separating red spirals from blue spirals.
Here, we use a combination of  \N  with the $(u-r)$ colour to separate early- from late-type galaxies. 
We specifically focus on the dependence of $(u-r)$ on \N[z][g] $= n_z/n_g$, but a similar result could be achieved by using other bands. 
The use of \N[z][g] reduces the contamination of late-type galaxies in the red-cloud and increases the number of recovered elliptical galaxies (see Tab.1). 
In addition, it provides information about the relative brightness and colour of the bulge and the disk in a late-type galaxy. 
Indeed, it correlates with the `lateness' of the spiral visual classification.  
It may therefore be useful for selecting samples of disk galaxies with specific properties, in lieu of bulge-disk decompositions. 
However, while the combination of \N with the $(u-r)$ colour creates a cleaner sample of elliptical galaxies, the $(u-r)$ and  $n_r$ combination is better on selecting late-type spirals, as we can see from Table~\ref{table:sample2}.

We stress the presence of some outlying spiral galaxies with high \N[z][g]  and slightly redder colour than the main bulk of spiral galaxies. 
All of those with \N[z][g]$>2$ have a bar.
In the presence of multi-components, e.g. bar or even disk, the single-\sersic function may be adversely affected. 
Fitting a galaxy with multi-component models can provide a more detail description of the galaxy but requires more assumptions regarding the structure of the galaxy.
The additional components are also more sensitive to the presence of spiral arms or other smaller substructures. 
The simplicity of single-\sersic fitting makes it a valuable approach for this study, especially for low S/N images where an accurate multi-component fit is not possible.
The fact that we can see these galaxies as outliers means that in general \N parameter help us to identify some of these cases.


In the lower left panel we show that using only three \sersic index measurements at three different wavelengths allows us to get a reasonable separation of early- from late-type galaxies.\footnote{In fact, a similar plot/result could be achieved by only using 2 \sersic indices, if one were to plot $n_g/n_r$ vs $n_r$  or $n_z/n_g$ vs $n_g$. 
Given the allowed linear variation of $n$ with wavelength, this is identical with a stretching of the y-axis. 
In this paper, we chose $n_r$ and $n_g/n_z$ in order to be consistent with V14 and the future \megamorph papers that will allow higher variation with wavelength for the $n$ parameters. 
While this is not important when a linear variation is used, it becomes important to have the longest possible wavelength range when higher freedom is used for the structural parameters.} 
This separation, however, is not as clear as in the previous two panels.
We find that a single \sersic index cut ($n_r>2.5$) in this panel selects 84\% of the elliptical galaxies and 25\% of the S0-Irr galaxies, and the classification cut  of two parameters (\N$<1.2$ and $n_r>2.5$) selects 68\% of the elliptical galaxies and 9\% of the S0-Irr galaxies. 
In contrast, the classification cut of the top right panel (\N$<1.2$ and $(u-r)>2.3$) selects 94\% and 7\% of the S0-Irr galaxies, creating a cleaner sample of elliptical galaxies.
In addition,  there seems to be a correlation between \N[z][g] and $n_{\rm r}$ for Sbc-Irr galaxies, in that spiral galaxies with small $n_{\rm r}$ tend to have higher \N[z][g] values. 
This means that galaxies with smaller optical \sersic index in $r$-band tend to change their \sersic index more drastically with wavelength.
However we do not see this trend in S0-Sb galaxies.


In the fourth panel we show a correlation between \N[z][g] and \R[z][g]$= \re(z)/\re(g)$ for early- and late-type galaxies, with both visually-classified types increasing their \N[z][g]  and \R[z][g]  quantities in a similar way (the Pearson linear correlation coefficient is r=0.61, 99\% significance).
However,  the change of the effective radius with wavelength is more subtle than that of the \sersic index for the E-S0 galaxies, with the result that these objects lay slightly apart from late-type galaxies (Pearson r=0.82, 99\% significance). 
Additionally, we note that E-S0s are placed on the 1:1 line.

The fourth panel is equivalent to Figure 18 in V14. The main difference compared to V14 is that we use the $z$-band in our analysis instead of the $H$-band. 
Our much more local galaxy sample shows significantly higher signal to noise properties, compared to V14, so we can see that there is less overlap between the blue and red cloud.

\subsection{The artificial redshifted sample}

We  now inspect the artificially redshifted data at $z=0.11$, in order to test the robustness of these methods at higher redshift and/or lower quality data.\footnote{We chose the highest possible redshift that contains all the galaxies used in the original sample. 
At higher redshift, some galaxies do not have artificially redshifted images due to resolution issues (see V13 for further details), so a comparison at higher redshift might show biases.}
Recall that, as described in Sec.~\ref{sec:data}, these images have been degraded in resolution and signal-to-noise to mimic the effects of observing the galaxies at a higher redshift.
However, for simplicity of interpretation, no band-shifting correction has been applied.
Figure~\ref{fig:MMz} generally shows that the technique is able to separate galaxies at redshifts where a visual classification becomes more difficult due to both data quality and volume \citep{tex:BN09}.

We use the same cut in colour,  \N[z][g]  and \R[z][g]   at $z=0.11$ as we used for the original images  (z$<$0.01). 
We do not observe any systematic change of the derived parameters due to redshifting effects.
General trends and separations found in Figure~\ref{fig:MMzg} can still be found in this artificially redshifted sample (see Fig. ~\ref{fig:MMz}).
However we see a significant increase of the scatter that occurs naturally due with lower signal-to-noise data.
While at low redshift  \N is able to recover the same or an even cleaner sample of elliptical galaxies by using colour and \sersic index, at higher redshifts both classifiers show a significant drop in their ability to separate the two populations. 

Especially in the third panel we notice that the scatter has increased for the early disky galaxies. 
We find that most of them have been moved below the \N=1.2 line creating a blob and loosing their distinct placement found in Figure~\ref{fig:MMzg}. 
Furthermore, as the image resolution decreases,  bars are becoming less and less visible and their presence does not affect the \sersic index measurements as much. 
As a result, most of the galaxies that have \N[z][g]$>2$ as z$<$0.01 have lower \N values at z=0.11.

\subsection{Discussion}


As discussed in V14, \N indicates how the central concentration of the light profile varies with wavelength, while \R quantifies the variation of galaxy size with wavelength. 
In Fig.~\ref{fig:MMzg} we find that our sample shows a wide range of \N values and small range of \R.
We suspect that \N varies significantly with Hubble type as it influenced by how dominant the bulge is in each galaxy and how this dominance changes with wavelength (i.e. the relative colour of bulge and disk).
The \sersic index will typically be higher in red bands because the bulge tends to be more dominant at longer wavelengths.
In galaxies with two similarly bright components with different colours, \N would be expected to depart most from unity.  
Indeed, most of the two-component galaxies in our sample that host different colour components (red bulge and blue disk; Sab-Sd) have \N$>$1.  
However, as shown in \citet{tex:HB13}, the effect of dust may partly explain values of \N, and so may have a role in its dependence on morphology.

In contrast to \N, we find that \R values are almost independent of their Hubble type. 
Although some differences were found in V14, they are less significant than those for \N.  
The narrower wavelength range we are considering here may further reduce the observed variation in \R.
In \citet{tex:VB14b} we showed that, for the sample considered in this paper, all galaxies, regardless of morphology, have a similar bulge-to-disk effective radius ratio ($R_{\rm e,b}/R_{\rm e,d}\sim$0.3), changing little with wavelength. 
However, it is currently unclear why the above arguments for \N do not translate into similar behaviour for \R.
We are continuing to explore the behaviour and physical interpretation of \N and \R, using the sample of V14 (Kennedy et al., in prep.).

Our classifications, based on \N, rely on contrasting radial colour variations in galaxies of different types. 
We have shown that, for our reasonably representative sample, these colour variations are sufficient to distinguish galaxy morphologies. 
Our approach is particularly helpful for separating ellipticals from early-spirals, Sa-Sb, which often have similar colour and \sersic index.

Most of the spirals in our sample are in the field. 
Disk-dominated galaxies are known to change their overall colour with environment, such that they become redder at higher densities (\citealt{tex:BN09, tex:CC13b}).  
Red spirals are particularly challenging for existing automated methods to distinguish from earlier-types. 
Overall reddening may be accompanied by a reduction in radial colour variation, perhaps due to a reduced colour contrast between bulge and disk.  
In that case, \N may lose its ability to differentiate between (red) spirals and earlier-types.  
One should therefore take care in applying the findings of this paper to a cluster sample.  
However, \N will still provide some additional insight over colour and \sersic index alone.

The difficulty of identifying disks in clusters also affects visual classifications. 
Reduced contrast in the spiral arms, and between the bulge and disk, often results in red spirals being classified as S0s, particularly when viewed close to edge-on \citep{tex:BN09}. 
Even more S0s are mistaken for ellipticals.  
Kinematics provide the most powerful solution \citep{tex:CC11c}.  
In their absence, the wavelength-dependence of structure, via both single-component fits and decompositions (e.g. \citealt{tex:VB14b}), provides an additional tool with which to tackle the problem.

\section{Conclusions}  
\label{sec:conclusions}

In the context of the \megamorph project we have demonstrated that the change of \sersic index with wavelength (\N) can be used as a proxy for galaxy morphology. 
We have used galaxy structural parameters measured with a new multi-wavelength technique that allows us to fit the galaxy light profile with the use of multiple images at the same time as described in V13. 
This method of classifying galaxy morphology has been initially suggested in V14 but the sample used in Vulcani et al. had no visual morphological classifications in order to test the validity of this proxy.

In this paper, by considering a nearby sample of galaxies with visual morphological classifications, we find that the \N can be used as a galaxy classifier to separate early- from late-type galaxies. 
In order to investigate the ability of \N to select a pure sample of early-types, we combine it initially with the overall $(u-r)$ galaxy colour and then with the change of effective radius with wavelength (\R). 
By comparing these classifiers with a widely used classifier i.e. colour and \sersic index, we find that with the use of \N - $(u-r)$ returns better result both from the $(u-r)$ - $n$ and  \N - \R galaxy type separation. 
Roughly half of the spiral galaxies that contaminate an early-type sample selected using (u-r) - n are identified as late-types by \N - (u-r).

Regarding the \R classifier we find that the average sizes of both early- and late-type galaxies do not change substantially over optical wavelengths.

Additionally, we find that \N correlates with \R and the correlation is becoming stronger if we only consider early-type (E, S0) galaxies.

Using a sample of artificially-redshifted images, we demonstrate that this technique remains effective up to $z\sim 0.11$ for the resolution of our images. 
Nonetheless, we notice a significant increase of the scatter due to the increase of noise.

\section*{Acknowledgments}

This publication was made possible by NPRP grant \# 08-643-1-112 from the Qatar National Research Fund (a member of Qatar Foundation). The statements made herein are solely the responsibility of the authors.  BH and MV are supported by this NPRP grant.  SPB gratefully acknowledges an STFC Advanced Fellowship. BV was supported by the World Premier International Research Center Initiative (WPI), MEXT, Japan and by the Kakenhi Grant-in-Aid for Young Scientists (B)(26870140) from the Japan Society for the Promotion of Science (JSPS).  We would like to thank Carnegie Mellon University in Qatar and The University of Nottingham for their hospitality.  We also thank the referee for helpful suggestions and clarifications. 

Funding for the SDSS and SDSS-II has been provided by the Alfred P. Sloan Foundation, the Participating Institutions, the National Science Foundation, the U.S. Department of Energy, the National Aeronautics and Space Administration, the Japanese Monbukagakusho, the Max Planck Society, and the Higher Education Funding Council for England. The SDSS Web Site is http://www.sdss.org/. The SDSS is managed by the Astrophysical Research Consortium for the Participating Institutions.

\bibliography{references-mar3}

\onecolumn

\label{lastpage}

\end{document}